\documentclass[twocolumn,aps,rmp,floats]{revtex4}
\usepackage{graphicx}
\usepackage{amssymb}
\usepackage{epstopdf}
\usepackage{epsfig}
\begin{document}

\newcommand{\ignore}[1]{}

\begin{widetext}

\leftline{\Large\bf Information based clustering}
\bigskip

\leftline{\large Noam Slonim, Gurinder Singh Atwal, Ga\v{s}per Tka\v{c}ik, 
and William Bialek}
\bigskip
\noindent
Joseph Henry Laboratories of Physics, and\\
Lewis--Sigler Institute for Integrative Genomics\\
Princeton University, Princeton, New Jersey 08544\\
\{nslonim, gatwal, gtkacik, wbialek\}@princeton.edu
\bigskip

\begin{quote}
In an age of increasingly large data sets, investigators in many different disciplines 
have turned to clustering as a tool for data analysis and exploration.  
Existing clustering methods, however, typically depend on several nontrivial 
assumptions about the structure of data.
Here we reformulate the clustering problem from an information theoretic perspective 
which avoids many of these assumptions. 
In particular, our formulation obviates the need for defining a cluster "prototype," 
does not require an {\it a priori\/} similarity metric, 
is invariant to changes in the representation of the data,
and naturally captures non--linear relations. 
We apply this approach to different domains
and find that it consistently produces clusters that are more coherent than those
extracted by existing algorithms. 
Finally, our approach provides a way of clustering based on collective notions of similarity 
rather than the traditional pairwise measures.
\end{quote}
\end{widetext}

The idea that complex data can be grouped into clusters or categories
is central to our understanding of the world, and this structure arises 
in many diverse contexts        (e.g., Fig. 1).
In popular culture we group films or books into genres, in business we group companies 
into sectors of the economy, in biology we group the molecular components of cells 
into functional units or pathways, and so on. 
Typically these groupings are first constructed by hand using  
specific but qualitative                knowledge; 
e.g., Dell  and Apple belong in the same group because they both make computers.  
The challenge of clustering is to ask whether these        qualitative
groupings can be derived automatically from objective, quantitative data.  Is our intuition about sectors of the economy derivable, for example, from the dynamics of stock prices?  Are the functional units of the cell derivable from patterns of gene expression under different conditions (1,2)?
The literature on clustering, even in the
context of gene expression, is vast (3).
Our goal here is not to suggest yet another clustering algorithm, but
rather to focus on questions about the {\em formulation} of the
clustering problem. We are led to an approach, grounded in
information theory, that should have wide applicability. 

Our intuition about clustering starts with the obvious notion that
similar elements should fall within the same cluster while dissimilar ones
should not. But clustering also achieves data 
compression---instead of identifying each data point individually, we can identify
points by the cluster to which they belong, ending up with a simpler
and shorter description of the data.
Rate--distortion theory (4,5) formulates precisely the tradeoff between 
these two considerations, searching for assignments to clusters such that the number
of bits used to describe the data is minimized while 
the average similarity between each data point and its 
cluster representative (or prototype) is maximized. 
A well known limitation of this formulation (as in most approaches to clustering) 
is that one needs to specify the similarity measure in advance, 
and quite often this choice is made arbitrarily. Another issue, which attracts less attention, 
is that the notion of a representative or "cluster prototype" 
is inherent to this formulation although it is not always obvious how to define this concept.
Our approach provides plausible answers to both these concerns,
with further interesting consequences. 

\section*{Theory}

{\bf Theoretical Formulation}. Imagine that there are $N$ elements (${\rm i} = 1, 2, \cdots, N$)
and $N_c$ clusters ($C = 1, 2, \cdots, N_c$) and that 
we have assigned elements ${\rm i}$ to clusters $C$
according to some probabilistic rules, $P(C|{\rm i})$, 
that serve as the variables in our analysis.\footnote{Conventionally, one distinguishes ``hard'' clustering, in which each element is assigned to exactly one cluster, and ``soft'' clustering in which the assignments are probabilistic, described by a conditional distribution $P(C|{\rm i})$; we consider here the more general soft clustering with hard clustering emerging as a limiting case.}
If we reach into a cluster and pull out elements at random, we would like these elements
to be as similar to one another as possible.  
Similarity usually is defined among pairs of elements 
(e.g., the closeness of points in some metric space), 
but as noted below we also can construct more collective measures 
of similarity among $r > 2$ elements; 
perhaps surprisingly we will see that that this more general case can 
be analyzed at no extra cost. Leaving aside for the
moment the question of how to measure similarity, let us assume that computing the similarity 
among $r$ elements ${\rm i}_1, {\rm
i}_2, \cdots, {\rm i}_r$ returns a similarity measure $s({\rm i}_1,
{\rm i}_2, \cdots, {\rm i}_r)$. 
The average similarity among elements chosen 
out of a single cluster is
\begin{equation}
s(C) = \sum_{{\rm i}_1 =1}^N \cdots \sum_{{\rm
i}_r = 1}^N P({\rm i}_1 | C) \cdots P({\rm i}_r | C)
s({\rm i}_1, \cdots , {\rm i}_r ),
\end{equation}
where $P({\rm i}|C)$ is the probability to find element $\rm i$ in
cluster $C$. 
This average similarity corresponds to a scenario where
one chooses the elements $\{{\rm i}_1,\cdots,{\rm i}_r\}$ at random out of a cluster $C$,
independently of each other; other formulations might also be plausible. 
From Bayes' rule we have 
$P({\rm i}|C) = P(C|{\rm i})P({\rm i})/P(C)$, 
where $P(C)$ is the total probability of finding        any element in cluster $C$,
$P(C) = \sum_{\rm i} P(C|{\rm i}) P({\rm i})$. In many cases the elements ${\rm i}$ occur with equal probability so 
that $P({\rm i}) = 1/N$. 
We further consider this case for simplicity, although it is not essential. 
The intuition about the ``goodness''  of the clustering is 
expressed through the average similarity over all the clusters,
\begin{equation}
\langle s \rangle = \sum_{C =1}^{N_c} P(C) s(C).
\end{equation}
For the special case of pairwise ``hard'' clustering 
we obtain $\langle s \rangle_h = \frac{1}{N}\sum_{C,i,j}\frac{1}{|C|}s({\rm i},{\rm j})$,
where $|C|$ is the size of cluster $C$. This simpler form was shown in (6)
to satisfy basic invariance and robustness criteria.

The task then is  
to choose the assignment rules $P(C|{\rm i})$ that maximize $\langle s \rangle$, 
while, as in rate--distortion theory, 
simultaneously compressing our description of the data as much as possible.
To implement this intuition we maximize $\langle s \rangle$ while constraining the  
information carried by the cluster identities (5),
\begin{equation}
I (C; {\rm i}) = {1\over N} \sum_{{\rm i}=1}^N \sum_{C =1}^{N_c} 
P(C|{\rm i}) \log\left[{{P(C|{\rm i})}\over{P(C)}}\right] .
\end{equation}
Thus, our mathematical formulation of the intuitive clustering problem 
is to maximize the functional
\begin{equation}
{\cal F} = \langle s \rangle - T I(C;{\rm i}),
\label{F}
\end{equation}
where the Lagrange multiplier $T$ enforces the constraint on 
$I(C;{\rm i})$. Notice that, as in other formulations of the clustering problem,
 ${\cal F}$ resembles 
the free energy in statistical mechanics, where
the temperature $T$ specifies the tradeoff between energy and entropy
like terms. 

This formulation is intimately related to conventional rate--distortion theory.
In rate--distortion clustering one is given a fixed
number of bits with which to describe the data, 
and the goal is to use these bits
so as to minimize the distortion between the data elements and some 
representatives of these data. 
In practice the bits specify membership
in a cluster, and the representatives are prototypical or average
patterns in each cluster.  Here we see that we can formulate a similar tradeoff 
with no need to introduce the notion of a representative or average; 
instead, we measure directly the similarity of elements within each cluster;
moreover, we can consider collective rather than pairwise measures of similarity.
A more rigorous treatment detailing the relation between Eq. (4) and
the conventional rate--distortion functional will be presented elsewhere.\\

{\bf Optimal Solution}. In general it is not possible to find an explicit solution for the
$P(C|{\rm i})$ that maximize ${\cal F}$.  However, 
if we assume that ${\cal F}$ is differentiable with respect to the variables $P(C|{\rm i})$,
equating the derivative to zero yields after some algebra
a set of implicit, self--consistent equations that any optimal 
solution must obey:
\begin{equation}
P(C|{\rm i}) = {{P(C)} \over {Z({\rm i}; T)}}\exp\bigg{\{} {1\over T}[ r s(C;{\rm i} ) - (r-1) s(C)]\bigg{\}},
\label{selfcon}
\end{equation}
where $Z({\rm i}; T)$ is a normalization constant and
$s(C;{\rm i})$ is the 
expected similarity between 
$\rm i$ and $r-1$ members of cluster $C$,
\begin{eqnarray}
s(C;{\rm i}) &=& \sum_{{\rm i}_1 =1}^N \cdots \sum_{{\rm i}_{r-1} = 1}^N P({\rm i}_1 | C) \cdots\\
\nonumber
&~& P({\rm i}_{r-1} | C) s({\rm i}_1, \cdots , {\rm i}_{r-1}, {\rm i} ) .
\end{eqnarray}
The derivation of these equations from the optimization of ${\cal F}$ is
reminiscent of the derivation of the rate--distortion
(5) or information bottleneck (7) equations.
This simple form is valid when the similarity measure is invariant under permutations of the arguments. In the more general case we have
\begin{equation}
P(C|{\rm i}) = {{P(C)} \over {Z({\rm i}; T)}}\exp\bigg{\{} {1\over T}[ \sum_{r'=1}^r s(C;{\rm i^{(r')}} ) - (r-1) s(C)]\bigg{\}},
\end{equation}
where $s(C;{\rm i^{(r')}})$ is the expected similarity between ${\rm i}$ and $r-1$ members of cluster $C$ when ${\rm i}$ is the $r'$ argument of $s$.

An obvious feature of Eq. (5) is that element $\rm i$ should be assigned to
cluster $C$ with higher probability if it is more similar to the other
elements in the cluster. Less obvious is that this similarity has to
be weighed against the mean similarity among all the elements in the
cluster. Thus, our approach automatically embodies the intuitive
principle that ``tightly knit'' groups are more difficult to join.
We emphasize that we did not 
explicitly impose this property, 
but rather it emerges directly from the variational principle of
maximizing ${\cal F}$; 
most other clustering methods do not capture this intuition. 

The probability $P(C|{\rm i})$ in Eq. (5) has the form of a
Boltzmann distribution, and increasing similarity among elements of a
cluster plays the role of lowering the energy; the 
temperature $T$ sets the scale for converting similarity
differences into probabilities. 
As we lower this temperature there are a sequence of ``phase
transitions'' to solutions with more distinct
clusters that achieve greater mean similarity in each cluster (8). 
For a fixed number of clusters, reducing the temperature yields
more deterministic $P(C|{\rm i})$ assignments.\\

{\bf Algorithm}.
Although Eq. (5) is an implicit set of equations, we can turn this self--consistency condition into an iterative algorithm that finds an
explicit numerical solution for $P(C|{\rm i})$ that corresponds
to a (perhaps local) maximum of ${\cal F}$. Fig. 2 presents 
a pseudo-code of the algorithm for the case $r=2$. 
Extending the algorithm for the general case of more than
pairwise relations ($r>2$) is straightforward. 
In principle we repeat this procedure for different initializations
and choose the solution which maximizes ${\cal F} = \langle s \rangle - T I (C; {\rm i})$.
We refer to the algorithm described here as {\it Iclust\/}.
We emphasize that we utilize this algorithm mainly because
it emerges directly out of the theoretical analysis. 
Other procedures that aim to optimize the same
target functional are certainly plausible and we expect future research 
to elucidate the potential (dis)advantages of the different alternatives.\\

{\bf Information as a Similarity Measure}.
In formulating the clustering problem as the optimization of ${\cal F}$,
we have used, as in rate--distortion theory,  
the generality of information theory to provide a natural measure
for the cost of dividing the data into more clusters,
but the similarity measure remains arbitrary and
commonly is  believed to be problem specific.  
Is it possible to use 
information theory to address this issue as well?
To be concrete, consider the case where the elements $\rm i$ are genes
and we are trying to measure the relation between gene expression
patterns across a variety of conditions $\mu = 1, 2 , \cdots , M$;
gene $\rm i$ has expression level $e_{\rm i}(\mu )$ under condition
$\mu$.  We imagine that there is some real distribution of conditions
that cells encounter during their lifetime, and an experiment with a
finite set of conditions provides samples out of this distribution.
Then, for each gene we can define the probability 
density of expression levels,
\begin{equation}
P_{\rm i}(e) = {1\over M}\sum_{\mu =1}^M \delta
(e-e_{\rm i}(\mu)) ,
\end{equation}
which should become smooth as $M\rightarrow \infty$.  Similarly we can
define the joint 
probability density  
for the expression levels of $r$ genes $i_1, i_2, \cdots, i_r$,
\begin{equation}
P_{\rm i_1 \cdots i_r}(e_1, \cdots , e_r) = {1\over M}\sum_{\mu =1}^M\delta (e_1-e_{\rm i_1}(\mu)) 
\cdots
\delta (e_r-e_{\rm i_r}(\mu)).
\end{equation}
Given the joint distributions of expression levels, information theory provides
natural measures of the relations among genes. 
For $r=2$, we can
identify the relatedness of genes $\rm i$ and $\rm j$ with the mutual
information between the expression levels,
\begin{eqnarray}
s({\rm i},{\rm j}) = I_{{\rm i},{\rm j}} &=& \int de_1 \int de_2\, P_{\rm ij} (e_1 , e_2) \cdots\\
\nonumber
&~& \log_2 \left[ {{P_{\rm ij}(e_1 , e_2 )}\over{P_{\rm i}(e_1) P_{\rm j}(e_2)}}   \right] \,{\rm bits}.
\end{eqnarray}
This measure is naturally extended to the
multi--information among multiple variables (9), or genes:
\begin{eqnarray}
I^{(r)}_{{\rm i}_1,{\rm i}_2 , \cdots , {\rm i}_r} &=& \int d^re\,   P_{\rm i_1 i_2 \cdots i_r}(e_1, e_2 , \cdots , e_r) \cdots \\
\nonumber
& ~ & \log_2 \left[ {{P_{\rm i_1 i_2 \cdots i_r}(e_1, e_2 , \cdots , e_r)}\over{P_{\rm i_1}(e_1) P_{\rm i_2}(e_2) \cdots P_{\rm i_r}(e_r)}}   \right] \,{\rm bits}.
\end{eqnarray}

We recall that the mutual information is the {\it unique\/} measure of
relatedness between a pair of variables that obeys several simple and
desirable requirements independent of assumptions about the form of
the underlying probability distributions (4).  
In particular, the mutual (and multi--) information are independent of 
invertible transformations on the individual variables. For example, the mutual
information between the expression levels of two genes is identical to
the mutual information between the log of the expression levels: there is no
need to find the ``right'' variables with which to represent the data.
The absolute scale of the information measure also has a clear meaning. 
For example, if two genes share more than one bit of information then
the underlying biological mechanisms must be more subtle than just
turning expression on and off. 
In addition, the mutual information 
reflects any type of dependence among variables while ordinary
correlation measures typically ignore nonlinear dependences. 

While these theoretical advantages are well known, in practice information theoretic 
quantities are notoriously difficult to estimate from finite data. 
For example, although the distributions in Eq's. (8,9) become smooth in the limit of many samples ($M \rightarrow \infty$), with a finite amount of data one needs to regularize or discretize the distributions, and this could introduce artifacts.  Although there is no completely general solution to these problems, we have found that in practice the difficulties are not as serious as one might have expected.  Using an adaptation of the ``direct'' estimation method originally developed in the analysis of neural coding (10), we have found that one can obtain reliable estimates of mutual (and sometimes multi--) information values for a variety of data types, 
including gene expression data (11); see the supplementary material for details. 
In particular, experiments which explore gene expression levels 
under $> 100$ conditions are sufficient to estimate the mutual information 
between pairs of genes with an accuracy of $\sim 0.1\,$ bits.\footnote{It should be noted that in applications where there is a natural similarity measure  it might be advantageous to use this measure directly. Furthermore, in situations where the number of observations is not sufficient for non--parametric estimates of the information relations, other heuristic similarity measures should be employed or one could use parametric models for the underlying distributions. Notice, though, that these alternative measures can be incorporated into the algorithm in Fig. 2.}

To summarize, we have suggested a purely information theoretic
approach to clustering and categorization: relatedness among elements
is defined by the mutual (or multi--) information, and optimal
clustering is defined as the best tradeoff between maximizing this
average relatedness within clusters and minimizing the number
of bits required to describe the data. 
The result is a formulation of clustering that trades bits of similarity 
against bits of descriptive power, with no further assumptions.
A freely available web implementation,
of the clustering algorithm and the mutual information estimation procedure 
can be found at {\it http://www.genomics.princeton.edu/biophysics-theory\/}.

\section*{Results}

{\bf Gene Expression}.
As a first test case we consider experiments on the response of gene expression levels in
yeast to various forms of environmental stress (12).  
Previous analysis identified a group of $\sim 300$ stress--induced and $\sim 600$
stress--repressed genes with 
{\it ``nearly identical but opposite patterns of expression in response to the environmental shifts''\/}
(13), and these genes were termed the environmental stress response (ESR) module. 
In fact, based on this observation, these genes were excluded from 
recent further analysis of the entire yeast genome 
(14). 
Nonetheless, as we shall see next, our approach automatically reveals
further rich and meaningful substructure in these data. 

As seen in Fig. 3A, differences in expression profiles
within the ESR module indeed are relatively subtle. 
However, when considering the mutual information relations (Fig. 3B) a relatively
clear structure emerges. 
We have solved our clustering problem for $r=2$ and various numbers of
clusters and temperatures.  The resulting concave tradeoff curves between 
$\langle s \rangle$ and 
$I(C;{\rm i})$ are shown in Fig. 4A.
We emphasize that  
we generate not a single solution, but a whole family of solutions 
describing structure at different levels of complexity.    
With the number of clusters fixed, $\langle s \rangle$  
gradually saturates as the temperature is lowered and the constraint
on $I(C;{\rm i})$ is relaxed. 
For the sake of
brevity we focused our analysis on the four solutions for
which the saturation of $\langle s \rangle$ is relatively clear ($1/T=25$).
At this temperature,        $\sim 85\%$ of the genes have nearly
deterministic assignments to one of the clusters
[$P(C|{\rm i}) > 0.9$ for a particular $C$].  
As an illustration, three of the twenty clusters found at this temperature 
are in fact the clusters presented in Fig. 1. 

We have assessed the biological significance of our results by
considering the distribution of gene annotations
across the clusters and estimating the corresponding clusters' 
{\it coherence\/} \footnote{Specifically, the coherence of a cluster (14) is defined as the percentage of elements in this cluster which are annotated by an annotation that was found to be significantly enriched in this cluster (P--val $< 0.05$, with the Bonferroni correction for multiple hypotheses). See the supplementary material for a detailed discussion regarding the statistical validation of our results.} with respect to all three Gene
Ontologies (15). 
Almost all of our clusters were significantly enriched in particular
annotations. We compared our performance to $18$ different conventional clustering
algorithms that are routinely applied to this data type (16). We employed the clustering software, available at {\it http://bonsai.ims.u-tokyo.ac.jp/~mdehoon/software/cluster/}, to implement the conventional algorithms. 
In Fig. 5 we see that our clusters obtained the highest average coherence,
typically by a significant margin. Moreover, even when the competing algorithms
cluster the $\log_2$ of expression (ratio) profiles---a common regularization 
used in this application with no formal 
justification---our results are comparable or superior to all of the alternatives. 

Instead of imposing a hierarchical structure on the data,
as done in many popular clustering algorithms,
here we directly examine the relations between solutions at different numbers 
of clusters that were found {\it independently\/}.\footnote{In standard agglomerative or hierarchical clustering one starts with the most detailed partition of singleton clusters and obtains new solutions through merging of clusters. Consequently, one must end up 
with a tree-like hierarchy of clustering partitions, 
regardless of whether the data structure actually supports this description.}
In Fig. 6 we see that an approximate hierarchy
emerges as a result rather than as an implicit assumption,
where some functional modules (e.g., the ``ribosome cluster'',
$C_{18}$) are better preserved than others. 

Our attention is drawn also to the cluster $C_7$, 
which is found  repeatedly at different numbers of clusters. 
Specifically, at the solution with $20$ clusters,
among the $114$ repressed genes in $C_7$, $69$ have an
uncharacterized molecular function; this level of
concentration has a probability of $\sim 10^{-15}$ to have arisen by chance.  
One might have suspected that almost every process in the cell has a
few components that have not been identified, and hence that as these
processes are regulated there would be a handful of unknown genes that
are regulated in concert with many genes of known function.  
At least for this cluster, our results indicate a different scenario where a significant
portion of tightly co--expressed genes remain uncharacterized to date.\\

{\bf Stock Prices}.
To emphasize the generality of our approach we 
consider a very different data set, the day--to--day fractional
changes in price of the stocks in the Standard and
Poor's (S \& P) 500 list (available at {\it http://www.standardandpoors.com\/}), 
during the 
trading days of 2003. 
We cluster these data exactly as in our analysis of gene expression data.
The resulting tradeoff curves are shown in Fig. 4B,
and again we focus on the four solutions where $\langle s \rangle$ already
saturates.

To determine the coherence of the ensuing clusters we used the Global Industry Classification Standard 
(GICS) (available at {\it http://wrds.wharton.upenn.edu\/})
which classifies companies at four different levels: 
sector, industry group, industry, and sub-industry. 
Thus each company is assigned four annotations, which are organized in a hierarchical tree,
somewhat similar to the Gene Ontology hierarchical annotation (15).

As before, our average coherence performance is comparable to or superior to all the other $18$
clustering algorithms we examined (Fig. 5). Almost all our clusters, at various levels of $N_c$, exhibit a surprisingly high degree of coherence with respect to the 
``functional labels'' that correspond to the different 
(sub) sectors of the economy. The four independent solutions, at $N_c=\{5,10,15,20\}$ and $1/T=35$, naturally form an approximate hierarchy (see Fig. 10 of Supporting Material).

We have analyzed in detail the results for $N_c=20$  and $1/T=35$ where selections from three of the derived clusters are shown in Fig. 1. Eight of the clusters are found to be perfectly (100\%) coherent, capturing subtle differences between industrial sectors. For example, two of the perfectly coherent clusters segregate companies into either investment banking and asset management (e.g., Merill Lynch) or commercial regional banks (e.g., PNC). Even in clusters with less than perfect coherence we are able to observe and explain relationships between intra-cluster companies above and beyond what the annotations may suggest. For example, one cluster is enriched with ``Hotel Resorts and Cruise Line'' companies at a coherence level of $30\%$. 
Nonetheless, the remaining companies in this cluster seem also to be tied with the tourism industry,
like the Walt Disney Co., banks which specialise in credit card issuing 
and so on.\\

{\bf Movie Ratings}.
Finally, we consider a third test case of yet another different
nature: movie ratings provided by more than $70,000$ viewers
(the EachMovie database, see \\{\it http://www.research.digital.com/SRC/eachmovie/\/}). 
Unlike the previous cases, the data here is already naturally quantized since only six possible ratings were permitted. 

We proceed as before to cluster the $500$ movies
that received the maximal number of votes. 
The resulting tradeoff curves are presented in Fig. 4C.
Few clusters are preserved amongst the solutions at different numbers of $N_c$, suggesting that a hierarchical structure may not be a natural representation of the data.
Cluster coherence was determined with respect to the genre labels provided in the database: action, animation, art-foreign, classic, comedy, drama, family, horror, romance, and thriller. Fig. 5 demonstrates that our results are superior to all the other $18$ standard clustering algorithms. 

We have analyzed in detail the results for $N_c=20$  and $1/T=40$ where, once again, selections from three of the derived clusters are shown in Fig. 1. The clusters indeed reflect the various genres, but also seem to capture subtle distinctions between sets of movies belonging to the same genre. For example, two of the clusters are both enriched in the action genre, but one group consists mainly of science-fiction movies and the other consists of movies in contemporary settings. 

Details of all three applications are given in a separate technical report, deposited on ArXiv as {\it http://arxiv.org/abs/q-bio.QM/0511042}.

\section*{Discussion}

Measuring the coherence of clusters corresponds to asking if the automatic, 
objective procedure embodied in our optimization principle does indeed recover 
the intuitive labeling constructed by human hands.  
Our success in recovering functional categories in 
different systems 
using exactly the same principle and practical algorithm is encouraging. 
It should be emphasized that our approach is {\em not} a model of each system 
and that there is no need for making data--dependent decisions
in the representation of the data, nor in the definition of similarity.  

Most clustering algorithms embody---perhaps implicitly---different models 
of the underlying statistical structure.\footnote{For example, the $K$--means algorithm corresponds to maximizing the likelihood of the data on the assumption that these are generated through a mixture
of spherical Gaussians.} %
In principle, more accurate models should lead to more meaningful clusters. However, the question of how to construct an accurate model obviously is 
quite involved, raising further issues that often are addressed arbitrarily 
before the cluster analysis begins. 
Moreover, as is clear from Fig. 5, an algorithm or model which is successful
in one data type might fail completely in a different domain; even in the context of
gene expression, successful analysis of data taken under one set of conditions
does not necessarily imply success in a different set of conditions, 
even for the same organism. 
Our use of information theory allows us to capture the 
relatedness of different patterns independent of assumptions about the nature 
of this relatedness. Correspondingly, we have a single approach which achieves 
high performance across different domains.

Finally, our approach can succeed where other methods would fail qualitatively.
Conventional algorithms search for linear or approximately linear relations
among the different variables, while our information theoretic approach
is responsive to any type of dependencies, including strongly nonlinear structures.
In addition, while the cluster analysis literature has focused thus far
on pairwise relations and similarity measures,
our approach sets a sound theoretical framework for analyzing complex data
based on higher order relations. 
Indeed, it was recently demonstrated, both in principle (17)
and in practice (18), that in some situations 
the data structure is obscured at the pairwise level, 
but clearly manifests itself only at higher levels. 
The question of how common such data are, as well as the associated computational
difficulties in analyzing such higher order relations, is yet to be explored. 

\begin{acknowledgments}
We thank O Elemento and E Segal for their help in connection with the analysis of the ESR data, and C Callan, D Botstein, N Friedman, R Schapire and S Tavazoie for their helpful comments on early versions of the manuscript.  This work was supported in part by the National Institutes of Health Grant P50 GM071508.  GT was supported in part by the Burroughs--Wellcome Graduate Training Program in Biological Dynamics.
\end{acknowledgments}


\newpage
\begin{widetext}

\vspace{1in}
\begin{figure}
\begin{center}
\begin{tabular}{c}
    \psfig{figure=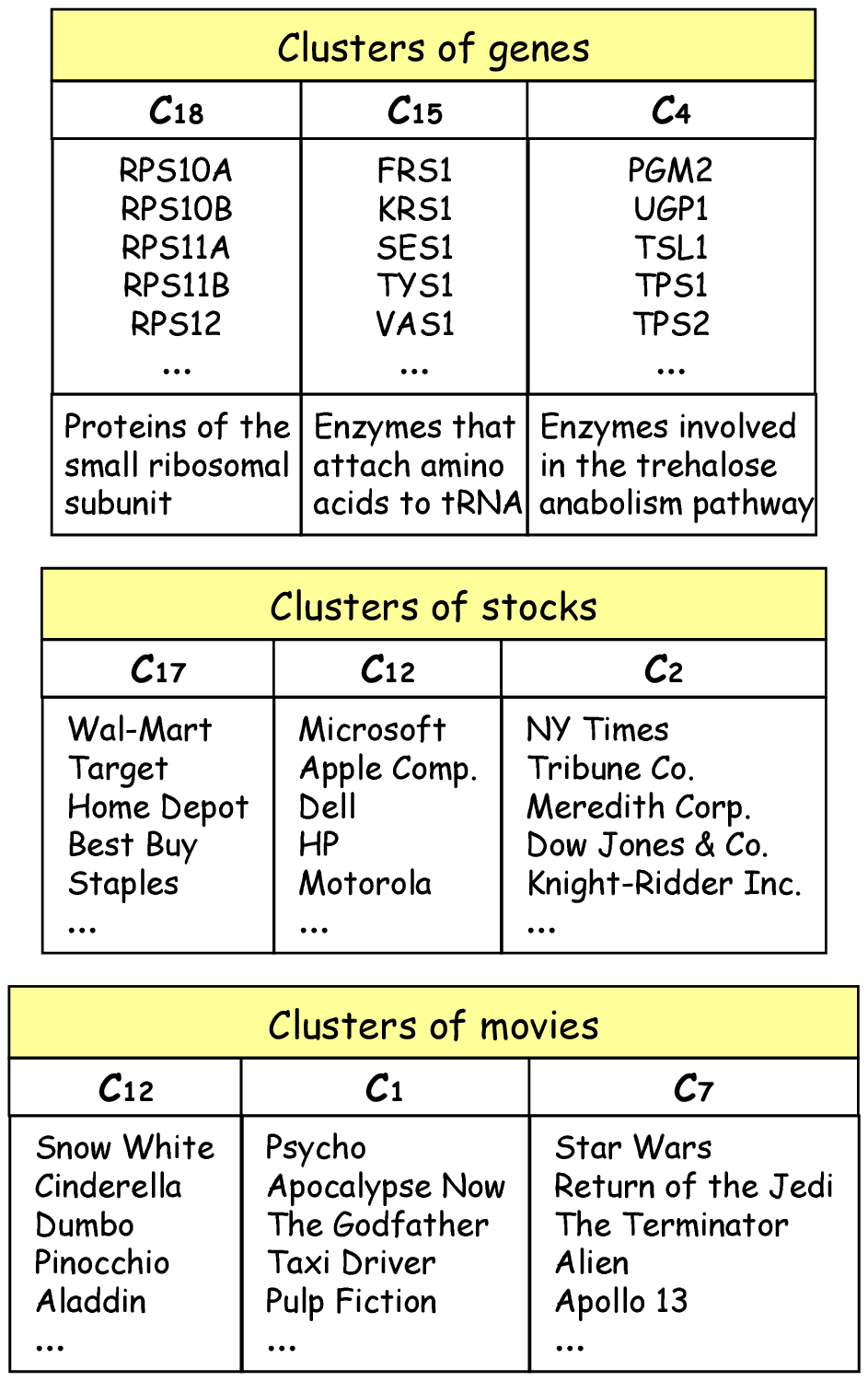,width=.7\columnwidth,height=6.0in} 
\end{tabular}
\caption{Examples of clusters in three different data sets.
For each cluster, a sample of five typical items is presented.
All clusters were found through the same automatic procedure.}
\end{center}
\end{figure}

\newpage

\ignore{
\vspace{1in}
\begin{figure}
\begin{center}
\begin{tabular}{c}
    \psfig{figure=Iclust_fig.eps,width=.7\columnwidth,height=6.0in} 
\end{tabular}
\caption{\small Pseudo-code of the iterative algorithm for the case of pairwise relations ($r=2$).}
\end{center}
\end{figure}
}

\begin{figure}
\centerline{\fbox{
\begin{minipage}{\columnwidth} 
\small
\begin{tabbing}
xxx\=xxx\=xxx\=xxx\=xxx\=xxx\=xxx\=\kill
\underline{\large\bf Input:}\\\\
\>Pairwise similarity matrix, $s(i_1,i_2),\;\forall\;i_1=1,...,N,\;i_2=1,...,N\;$.\\
\>Trade-off parameter, $T\;$.\\
\>Requested number of clusters, $N_c\;$.\\
\>Convergence parameter, $\epsilon\;$.\\\\
\underline{\large\bf Output:}\\\\
\>A (typically ``soft'') partition of the $N$ elements into $N_c$ clusters.\\\\
\underline{\large\bf Initialization:}\\\\
\>$m = 0\;$.\\
\>$P^{(m)}(C|i) \leftarrow\;$ A random (normalized) distribution $\;\forall\;i=1,...,N\;$.\\\\
\underline{\large\bf While True}\\\\
\>For every $i=1,...,N\;$:\\\\
\>\>$\bullet~~P^{(m+1)}(C|{\rm i}) \leftarrow {P^{(m)}(C)} \exp\bigg{\{} {1\over T}[ 2 s^{(m)}(C;{\rm i} ) - s^{(m)}(C)]\bigg{\}},\;\forall\;C=1,...,N_c\;$.\\\\
\>\>$\bullet~~P^{(m+1)}(C|{\rm i}) \leftarrow \frac{P^{(m+1)}(C|{\rm i})}{\sum_{C'=1}^{N_c} P^{(m+1)}(C'|{\rm i})},\;\forall\;C=1,...,N_c\;$.\\\\
\>\>$\bullet~~m \leftarrow m+1\;$.\\\\
\>If $\;\forall\;i=1,...,N,\;\forall\;C=1,...,N_c$ we have
$|P^{(m+1)}(C|i)-P^{(m)}(C|i)| \leq \epsilon\;$,\\
\>\>Break.\\\\
\end{tabbing}
\end{minipage}}}
\begin{center}
\caption{\small Pseudo-code of the iterative algorithm for the case of pairwise relations ($r=2$).}
\label{algorithm} 
\end{center}
\end{figure}

%
\begin{figure}[h] 
\begin{center}
\begin{tabular}{cc}
    \psfig{figure=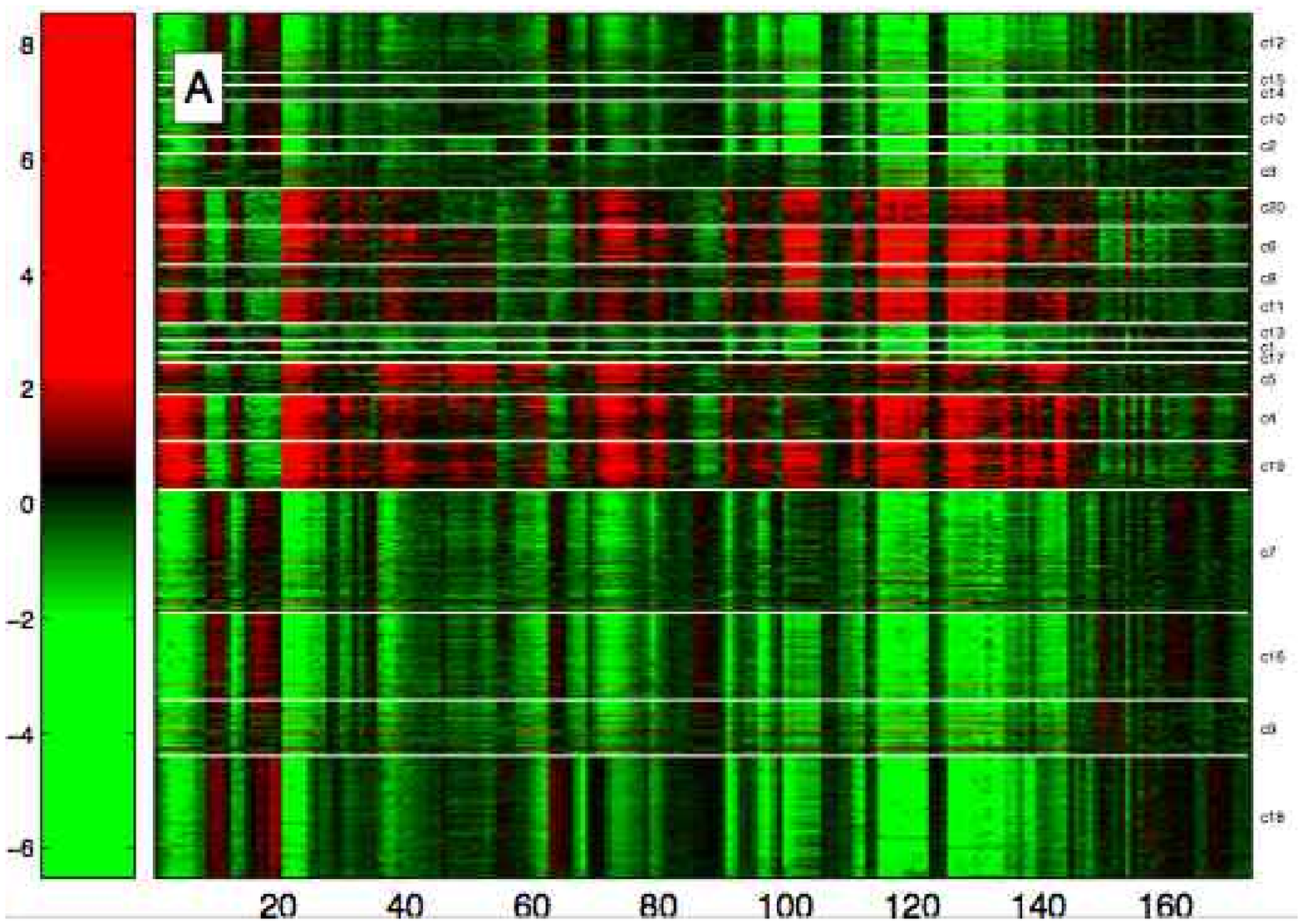,width=.5\columnwidth,height=4.0in} 
    \psfig{figure=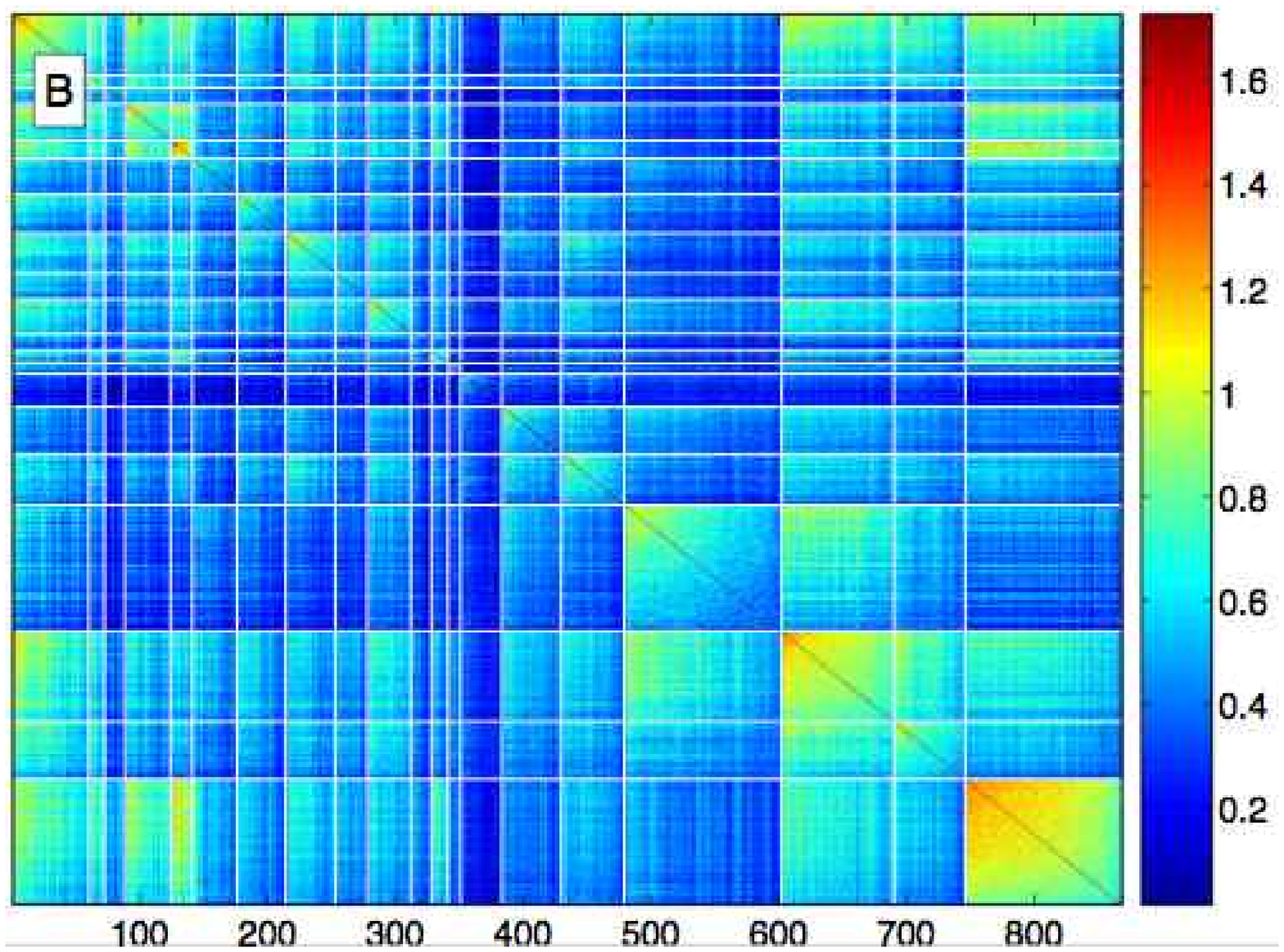,width=.5\columnwidth,height=4.0in} 
\end{tabular}
\end{center}
\caption{ESR data and information relations.
{\bf (A)} Expression profiles of the $\sim 900$ genes in the yeast ESR module
across the $173$ microarray stress experiments (12).
{\bf (B)} Mutual information relations (in bits) among the ESR genes.
In both panels the genes are sorted according to 
the solution with $20$ clusters and a relatively saturated $\langle s \rangle$.
Inside each cluster, genes are sorted according 
to their average mutual information relation with other cluster members.}
\label{ESR_MIs}
\end{figure}

\newpage
\begin{figure}[h] 
\label{Fig:TradeOffcurve}
\begin{center}
\begin{tabular}{ccc}
    \psfig{figure=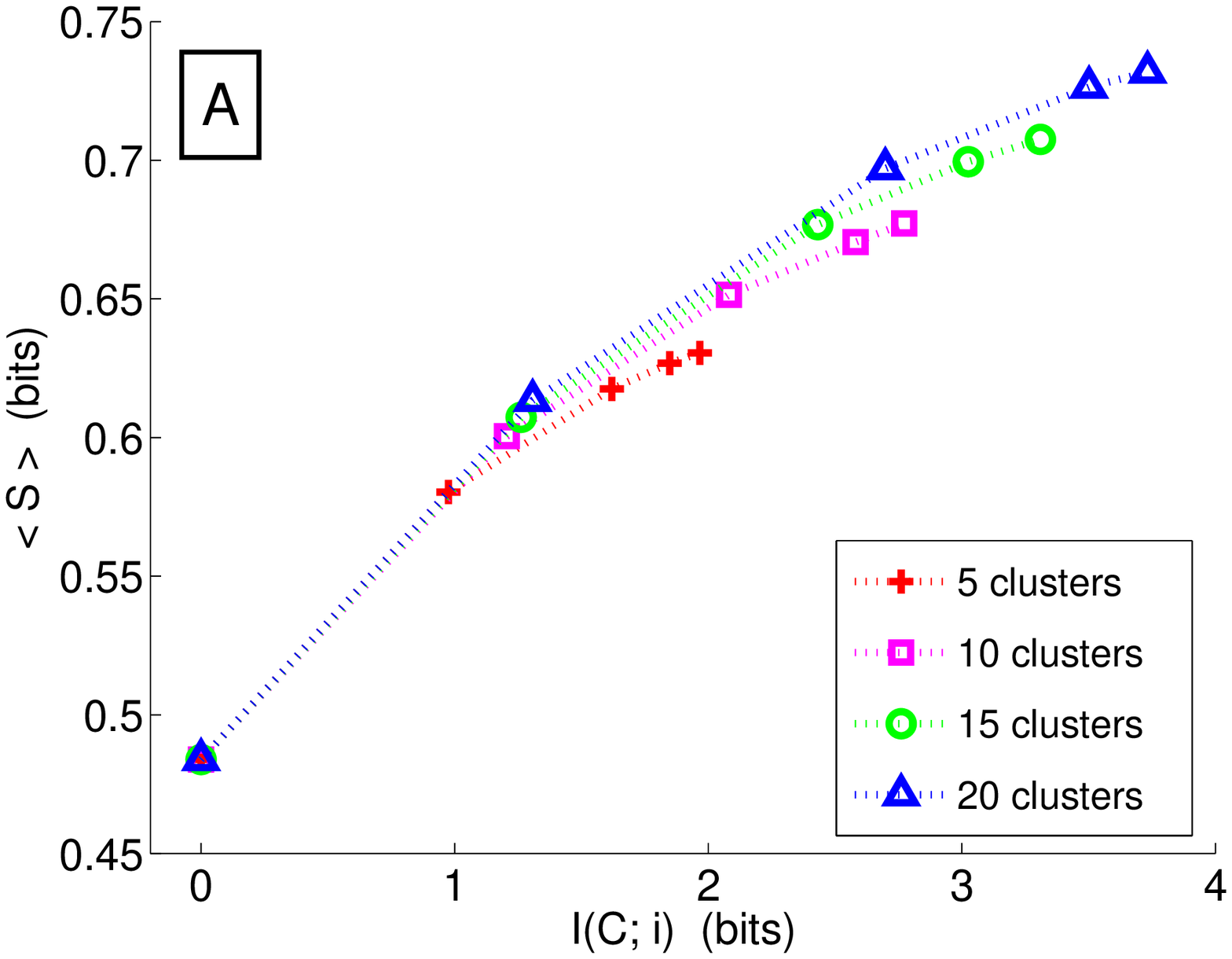,width=.31\columnwidth,height=2.0in} &
    \psfig{figure=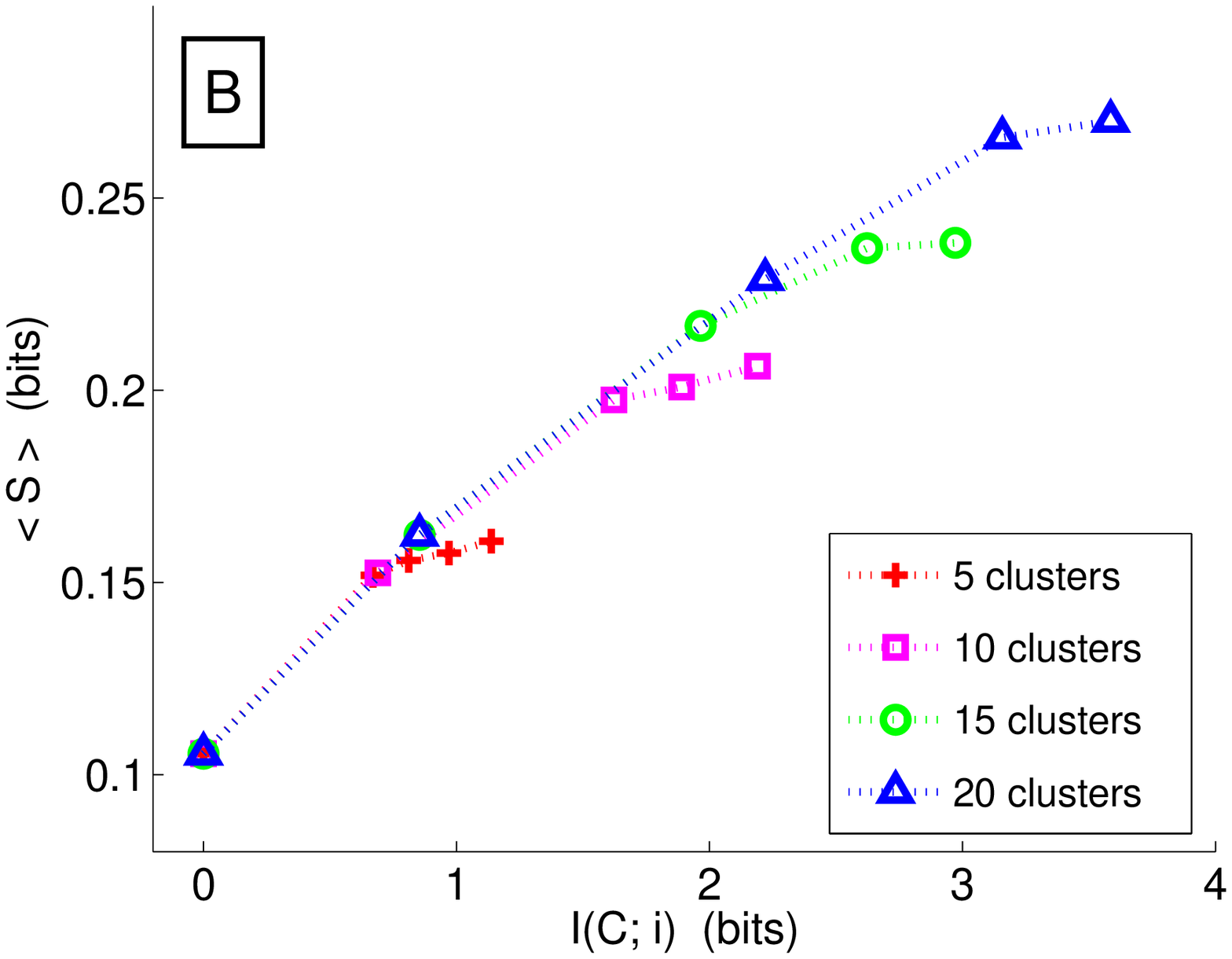,width=.31\columnwidth,height=2.0in} &
    \psfig{figure=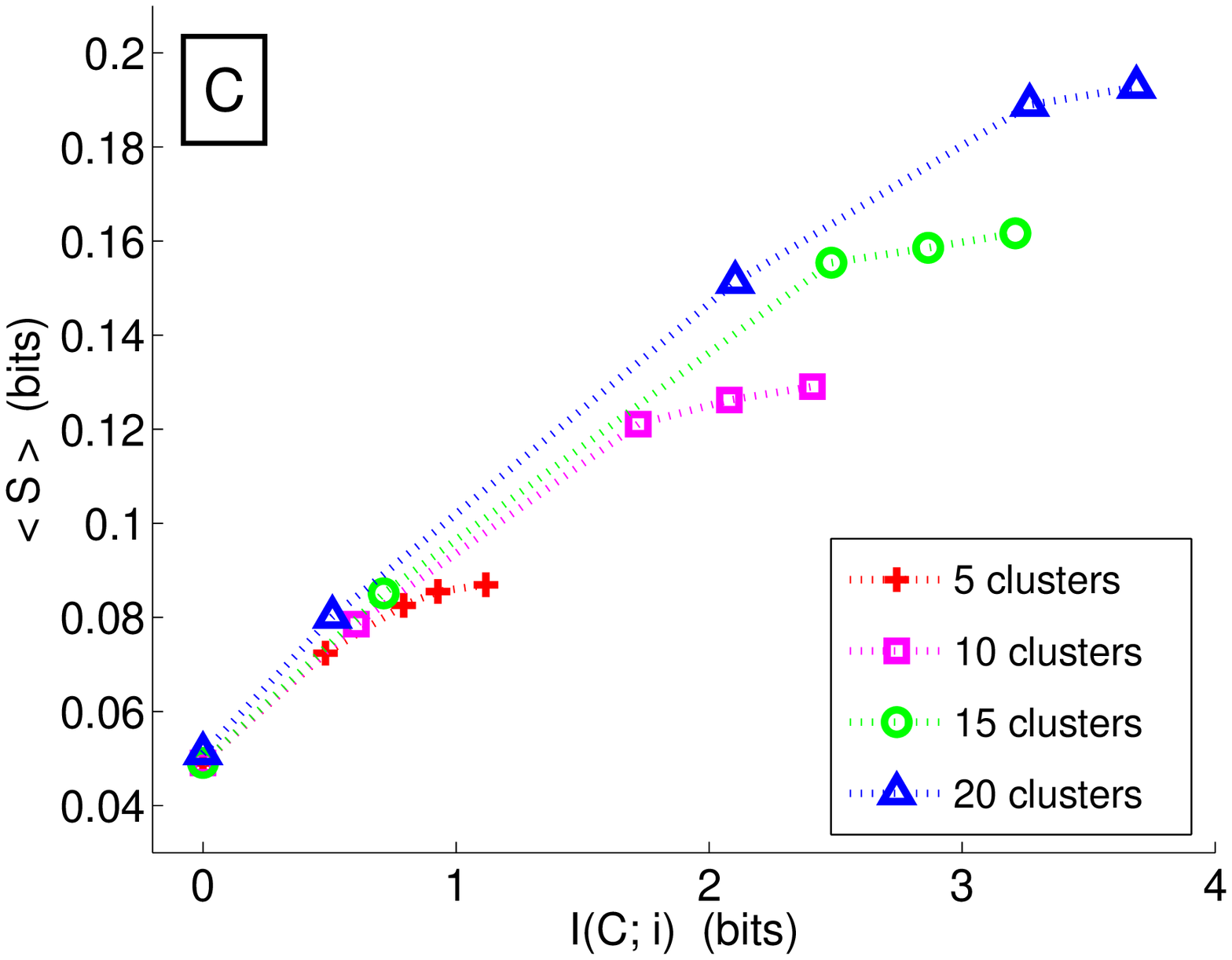,width=.31\columnwidth,height=2.0in} 
\end{tabular}
\end{center}
\caption{Tradeoff curves in all three applications.
In every panel, each curve describes the solutions obtained 
for a particular number of clusters. 
Different points along each curve correspond to different local maxima
of ${\cal F}$ at different $T$ values. 
{\bf (A)} Tradeoff curves for the ESR data with 
$\frac{1}{T}=\{5,10,15,20,25\}$.
In Fig. 6 we explore the possible hierarchical relations between 
the four saturated solutions at $\frac{1}{T}=25$.
{\bf (B)} Tradeoff curves for the S\&P 500 data with 
$\frac{1}{T}=\{15,20,25,30,35\}$.
{\bf (C)} Tradeoff curves for the EachMovie data with 
$\frac{1}{T}=\{20,25,30,35,40\}$.}
\end{figure}

\newpage
\begin{figure}[h] 
\begin{center}
\begin{tabular}{c}
    \psfig{figure=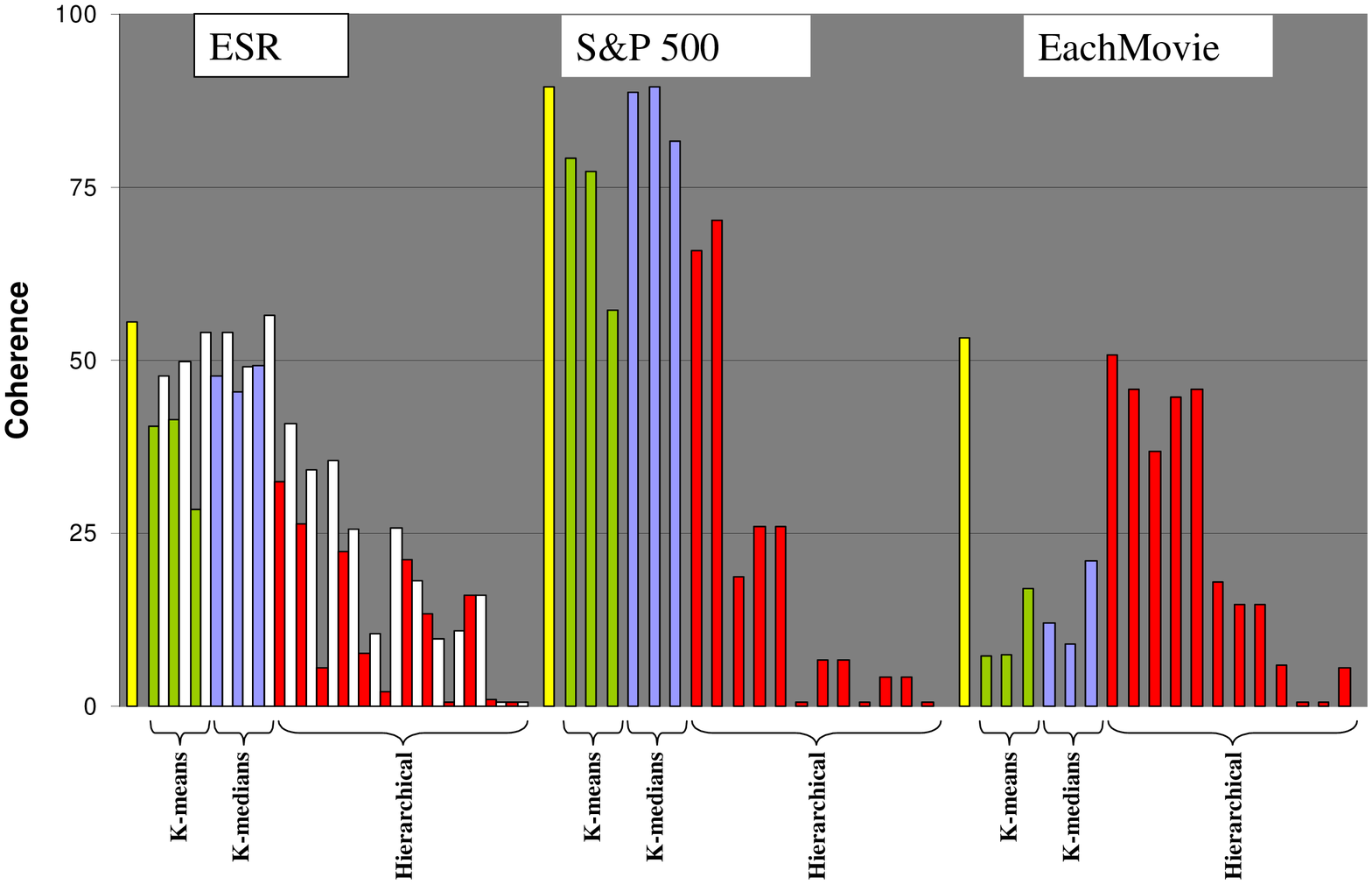,width=.95\columnwidth,height=5.0in} 
\end{tabular}
\end{center}
\caption{Comparison of coherence results of our approach (yellow) with 
conventional clustering algorithms (16):
$K$--means (green); $K$--medians (blue); Hierarchical (red).
For the hierarchical algorithms, four different variants are tried:
complete, average, centroid, and single linkage, respectively
from left to right. 
For every algorithm, three different similarity measures are applied:
Pearson correlation (left); absolute value of Pearson correlation
(middle); Euclidean distance (right).
The white bars in the ESR data correspond to applying the algorithm
to the $\log_2$ transformation of the expression ratios. 
In all cases, the results are averaged over all the different numbers of clusters that 
we tried: $N_c=5,10,15,20$. 
For the ESR data coherence is measured with respect to each of the three
Gene Ontologies and the results are averaged.}
\label{Fig:AvgCoh}
\end{figure}

\newpage
\begin{figure}[h] 
\begin{center}
\begin{tabular}{c}
    \psfig{figure=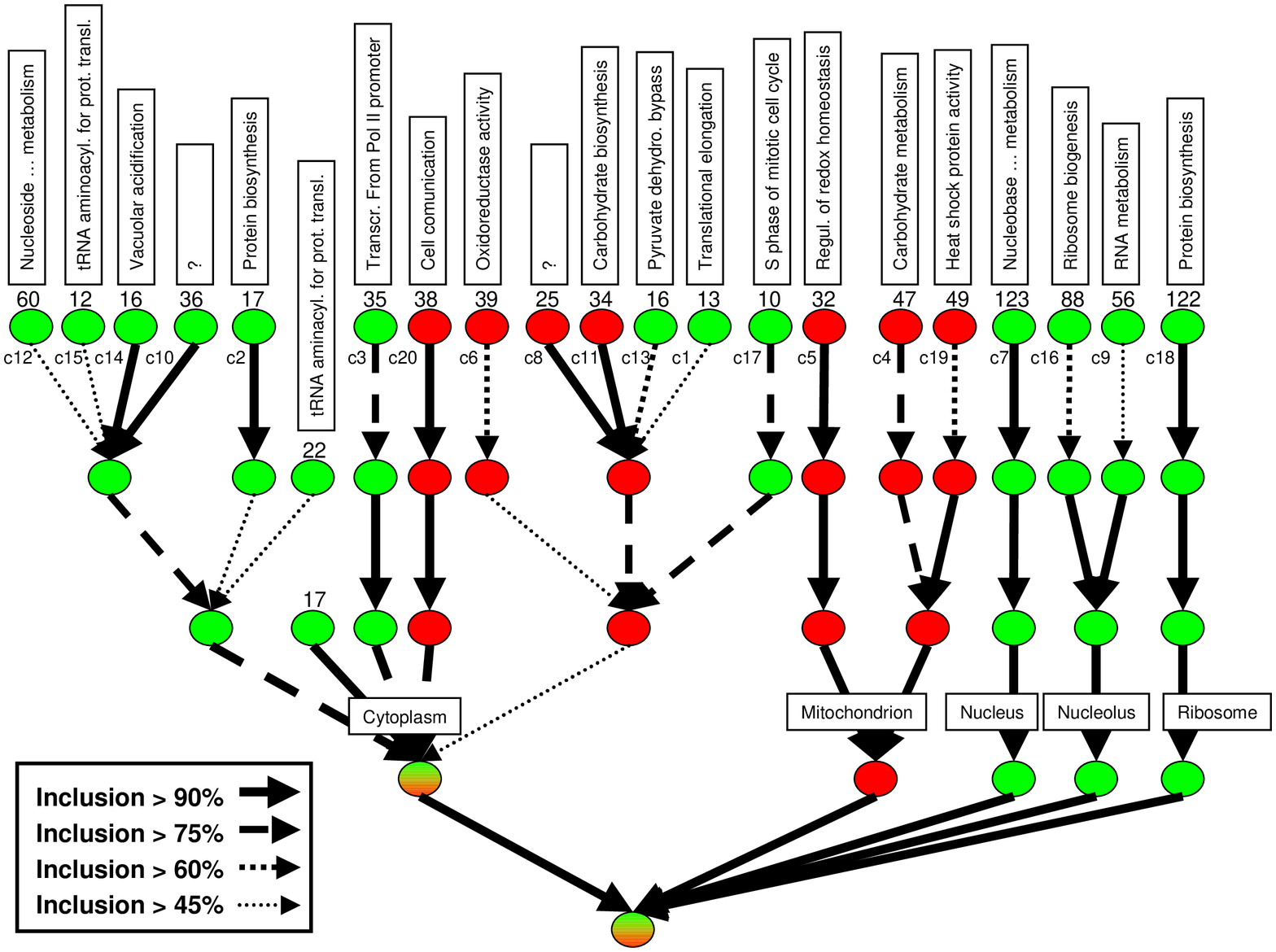,width=.9\columnwidth,height=5.5in} 
\end{tabular}
\end{center}
\caption{Relations between the optimal solutions with $N_c=\{5,10,15,20\}$
at $\frac{1}{T}=25$ for the ESR data.
Every cluster is connected to the cluster in the next 
-- less detailed -- partition that absorbs its most significant portion.
The edge type indicates the level of inclusion.
The independent solutions form an approximated hierarchical structure.
At the upper level 
the clusters are sorted as in Fig. 3.
The number above every cluster indicates the number of genes in it, 
and the text title corresponds to the most enriched GO
biological--process annotation in this cluster.
The titles of the five clusters at the lower level are 
their most enriched GO cellular-component annotation.
Most clusters were enriched with more than
one annotation, hence the short titles sometimes are too concise. 
Red and green clusters represent clusters with a clear majority of 
stress--induced or stress--repressed genes, respectively.}
\label{Fig:ESR_Hierarchy}
\end{figure}

\end{widetext}

\end{document}